\documentclass[sn-mathphys-ay]{sn-jnl}

\usepackage{multicol}%
\usepackage{graphicx}%
\usepackage{multirow}%
\usepackage{amsmath,amssymb,amsfonts}%
\usepackage{amsthm}%
\usepackage{mathrsfs}%
\usepackage[title]{appendix}%
\usepackage{xcolor}%
\usepackage{textcomp}%
\usepackage{manyfoot}%
\usepackage{booktabs}%
\usepackage{algorithm}%
\usepackage{algorithmicx}%
\usepackage{algpseudocode}%
\usepackage{listings}%

\theoremstyle{thmstyleone}%
\theoremstyle{thmstyletwo}%

\theoremstyle{thmstylethree}%

\usepackage{tikz}
\usetikzlibrary{arrows.meta, bending, decorations.markings}
\begin{document}
\title[The Dynamical Environment of {\sl Kepler--223}]{The Dynamical Environment of {\sl Kepler--223} Planetary System}


\author*[1]{\fnm{Carlos E.} \sur{Chavez}}\email{Carlos.ChavezPch@uanl.edu.mx}

\author[2]{\fnm{Adri\'an} \sur{Fierro}}
\equalcont{These authors contributed equally to this work.}

\author[2]{\fnm{\'Andres} \sur{Avil\'es}}
\equalcont{These authors contributed equally to this work.}

\affil*[1]{\orgdiv{Universidad Auton\'oma de Nuevo Le\'on}, \orgname{Facultad de Ingenier\'{i}a Mec\'anica y El\'ectrica}, \orgaddress{\city{San Nicol\'as de los Garza}, \postcode{66451}, \state{NL}, \country{M\'exico}}}

\affil[2]{\orgdiv{Universidad Auton\'oma de Nuevo Le\'on}, \orgname{Facultad de Ciencias F\'{i}sico--Matem\'aticas}, \orgaddress{\city{San Nicol\'as de los Garza}, \postcode{66451}, \state{NL}, \country{M\'exico}}}


\abstract{
The Kepler space telescope was a successful mission that used the transit method. Multi--planetary systems are interesting because they are dynamically rich due to the planets' interactions. Kepler's mission was the first to find a system with four planets all in resonance, that is  {\sl Kepler--223}.
The {\sl Kepler--223} system is of unique dynamical interest as the first exoplanetary system confirmed to host four sub--Neptune planets in a precise 3:4:6:8 resonant chain. Previous dynamical studies have struggled to fully recover the long--term phase protection mechanisms, often finding that higher--order resonant angles circulate rather than librate. In this work, we re--examine the dynamical state of {\sl Kepler--223} by performing high--resolution N--body integrations and constructing detailed stability maps in the $(a,e)$ phase space, we demonstrate that the system is locked in a deeper resonant state than previously thought. Crucially, we find that all resonant arguments, exhibit robust libration over secular timescales. Our stability maps reveal that the planets reside in clearly defined stability islands that allow for higher eccentricities than previous estimates. We further compare these numerical islands with analytical resonance widths and find excellent agreement for the inner three planets, while the outermost planet ({\sl Kepler--223}e) exhibits a stability region wider than first--order analytical theories predict, suggesting complex multi--body stabilization effects at the edge of the chain. \footnote{Published in Astrophysics and Space Science, Volume 371, Issue 5, id.46,  May 2026}} 



\maketitle

\section{Introduction}\label{sec1}

NASA’s Kepler mission is one of the most successful space--based exoplanet surveys to date\footnote{https://science.nasa.gov/mission/kepler/}. Over its nine years of operation, the Kepler space telescope detected approximately 2,700 exoplanets. Its discoveries have revolutionized our understanding of planetary system architectures, revealing that multi--planet systems are common, with at least 156 such systems identified and thousands of additional candidates still awaiting confirmation. The study of multi--planet systems is particularly compelling due to the complex gravitational interactions and rich dynamical behavior among their constituent planets. Of special interest are systems in mean--motion resonance (MMR), such as Kepler--223, whose present--day dynamical configurations preserve a fossil record of their formation and migration history.

Several exoplanetary systems have been identified in resonant configurations. Focusing on the Kepler catalog, notable examples include Kepler--29 \citep{bib7, bib24}, Kepler--31 \citep{bib1}, Kepler--36 \citep{bib1, bib2, bib29}, Kepler--37 \citep{bib36,bib35,bib30}, Kepler--51 \citep{bib23,bib20}, Kepler--60 \citep{bib9,bib15}, Kepler--80 \citep{bib21,bib31,bib22} and Kepler--88 \citep{bib37}, Kepler--223 \citep{bib25,bib12, bib26, bib32}. In addition, several systems are found near resonance, including Kepler--90 \citep{bib10,bib17, bib18}, Kepler--1649  \citep{bib14}, among others. After the conclusion of the primary Kepler mission, the K2 mission further expanded the sample by discovering resonant and near--resonant systems such as K2--32 and K2--138. Among all these systems, only Kepler--51, Kepler--60, Kepler--80, and Kepler--223 exhibit configurations in which all planets are locked in resonances. Notably, Kepler--223—the focus of this study—was the first system discovered in which all planets are fully engaged in a resonant chain.

Kepler--223 was continuously monitored by the Kepler space telescope during the entire primary mission, from March 2009 to May 2013 (approximately four years).

This system, with all planets near resonance, exhibits significant transit timing variations (hereafter TTVs), which provide strong constraints on the system parameters. In particular, the TTV analysis enabled the determination of planetary masses and indicated that all four planets must have low eccentricities in order for the system to remain dynamically stable. These results further suggest that the system is consistent with a migration--based formation scenario, a hypothesis that has been explored in subsequent studies.

\citet{bib25} analyzed the full Kepler data set for the Kepler--223 system and demonstrated that, based on the observational constraints, the system is indeed in resonance and dynamically stable over long timescales. Their analysis showed that resonance plays a crucial role in maintaining the stability of this compact system. In particular, all four sub--Neptune planets were found to be locked in a precise and long--lived 3:4:6:8 resonant chain.

Kepler--223 has subsequently been studied for a variety of reasons by several authors. \citet{bib12}, \citet{bib26}, \citet{bib42} and \citet{bib22} investigated the system in order to infer properties of the protoplanetary disk that formed the planets and likely shaped their final architecture. In addition, \citet{bib32} examined resonant chains and possible three--body resonances in Kepler--60 and Kepler--223, demonstrating that transit timing variation (TTV) signals can be misinterpreted as mean--motion resonance oscillations if not modeled dynamically, thereby highlighting the importance of self--consistent dynamical models.

However, characterizing the long-term stability of compact systems locked in deep resonances remains challenging. Recent large-scale studies have utilized fast statistical chaos indicators, such as the Mean Exponential Growth factor of Nearby Orbits (MEGNO) and the machine-learning algorithm SPOCK, to efficiently evaluate multiplanetary architectures \citep[e.g.,][]{bib40, bib52, bib39}. While computationally advantageous, these methods can be unreliable for systems deeply embedded in resonance. For instance, when applied to Kepler-223, these fast methods classify the system as rapidly chaotic, yielding a MEGNO value of $\approx 18$ \citep{bib52}. Given that Kepler-223 is an evolved system with an estimated age of 6.4 Gyr \citep{bib25}, it is dynamically highly unlikely that it has not achieved long-term stability. As noted by \citet{bib40} and \citet{bib39}, stability in such systems is often maintained by specific phase-protection processes that are easily misclassified as unstable by fast indicators. Therefore, full N-body numerical integrations---such as those performed in this work---are strictly necessary to properly capture the long-term phase-protection mechanisms present in highly resonant chains.

Consequently, determining the true robustness of the Kepler--223 resonant chain requires high--resolution, long--term N--body integrations capable of resolving the fine--scale structure of stability islands in phase space. Each multi--planet system in or near resonance occupies a distinctive region of phase space, typically described in terms of semi--major axis versus eccentricity. The topology of this phase space provides insight into the system’s stability, its proximity to stable resonant islands, and its likely formation history, as well as the extent of stable phase space available under external perturbations \citet{bib28}. 
\\
As Kepler--223 was the first planetary system discovered in a resonant chain, it provides an ideal laboratory for detailed dynamical studies. In this work, we adopt updated initial conditions kindly provided by Daniel Fabrycky, derived from the analysis of \citet{bib25} and \citet{bib32}, which allow us to model the system with higher fidelity than is possible using publicly available catalogs. Our primary objective is to map, at a higher resolution than in previous studies, the chaotic and stable regions of phase space surrounding each planet. By analyzing this real system—where all four planets interact strongly—we aim to identify the specific properties of the stability islands that ensure its long--term survival and to compare them with those of other resonant systems. This approach enables us to quantify the extent of phase--space protection for each planet and to determine whether they reside deep within stable regions or near chaotic boundaries, distinctions that are often overlooked by broader statistical surveys.

 This paper is organized as follows. In Section 1, we introduce the Kepler--223 system. Section 2 summarizes the system parameters, while Section 3 describes the numerical setup adopted in this work. In Section 4, we present the dynamical analysis and methodology, including the computation of resonant angles and the construction of stability maps. Our main results are presented in Section 5, and Section 6 concludes the paper with a discussion of the results and final remarks.


\section{The {\sl Kepler--223} System}
\label{Sec:Kepler223info}

Kepler--223 is a compact resonant--chain planetary system composed of four planets identified to date \citep{bib25}, orbiting a moderately evolved Sun--like star with an estimated age of approximately 6 Gyr. Early analyses of the Kepler photometric data initially suggested that two of the planets might occupy Trojan orbits \citep{bib19}. However, this interpretation was later revised when transit timing variations (TTVs) were taken into account, leading to the rejection of the Trojan configuration and the identification of the correct orbital architecture.

\citet{bib25} performed an extensive analysis of the transit and TTV data, which led to the discovery of the first exoplanetary system in which all four planets are locked in a 3:4:6:8 mean--motion resonant chain. The TTV analysis also enabled robust mass estimates for the planets, revealing that all four are sub--Neptune--sized objects.

These authors suggested that the long--term survival of the resonant configuration is partly due to the relatively large orbital separation of the innermost planet from the host star compared to other Kepler systems near resonance, such as Kepler--80. As a consequence, tidal dissipation effects are expected to be weak and insufficient to disrupt the resonant chain.

Given the advanced age of the system ($\sim$ 6 Gyr), Kepler--223 has had ample time to escape resonance if the configuration were unstable. Its persistence in a resonant state therefore provides strong constraints on its formation history and long--term dynamical stability.

\section{Initial Parameters}
\label{Sec:Initial}

Using spectroscopic observations, \citet{bib25} determined the fundamental physical properties of the host star. Kepler--223 is a solar--type star with a mass slightly larger than that of the Sun (1.125 $M_{\odot}$) and an estimated age of approximately 6 Gyr, indicating a moderately evolved evolutionary state

The planetary system properties were derived using the transit method applied to photometric data obtained by the Kepler space telescope (hereafter Kepler). These constraints were significantly refined through the analysis of transit timing variations (TTVs) observed over the full duration of Kepler’s primary mission, from March 2009 to May 2013 (approximately four years). The TTV signal in Kepler--223 is particularly strong due to the mutual gravitational interactions arising from the resonant coupling among all four planets, allowing robust constraints on both planetary masses and orbital eccentricities.

To infer the most likely system parameters, \citet{bib25} employed a Differential Evolution Markov Chain Monte Carlo approach (DEMCMC; \citet{bib33}). In their model, each planet was described by seven free parameters, while the host star was characterized by five parameters. The orbital evolution of each sampled configuration was numerically integrated using the MERCURY N--body package \citet{bib4}.

An initial constraint (hereafter $C_1$) was applied to discard solutions that, while providing good fits to the observational data, resulted in orbit crossing between adjacent planets. A second constraint ($C_2$) was then imposed to eliminate dynamically unstable solutions: 5,000 configurations that satisfied C1 were randomly selected and integrated for 10$^6$ years, retaining only those that remained stable over this timescale (2,008 solutions in total).

A final constraint ($C_3$) was applied by integrating the $C_2$ solutions over a shorter timescale of 100 years and requiring that all four planets remain locked in resonance. Only configurations satisfying this criterion were retained as the final parameter set.

Based on this analysis, we adopt the best--fit initial conditions listed in Extended Data Table 3 of \citet{bib25}, corresponding to the $C_3$ solutions (bottom values in the table), to ensure long--term stability. In addition, updated initial conditions obtained using DEMCMC combined with gas--driven migration were kindly provided by Daniel Fabrycky. After testing several configurations, we find that the $C_3$ solution provides an excellent starting point for our study, requiring only minor adjustments to the arguments of periapsis ($\omega$) and mean anomalies ($M$) to place the system deep within the most stable region of the resonant phase space. The final initial conditions adopted in this work are summarized in Table 1.

We adopt the standard coordinate system defined by  \citet{bib45}, in which the observer is located along the z--axis and the mean anomaly is measured relative to the Laplace vector (periapsis). \citet{bib25} also explored the inclusion of post--Newtonian (general relativistic) corrections in the MERCURY integrator to account for relativistic precession. Their tests showed that the inclusion of these terms did not significantly affect the results across 100 trials, and we therefore neglect relativistic corrections in the present work.

In the following sections, we describe our methodology for characterizing the dynamical environment surrounding each planet.

In all the following sections we explain how we study the dynamic environment around each planet.

\subsection{Kepler--223 and the four known planets}

The star {\sl Kepler--223}, also known as KOI--30, KIC 10227020, and 2MASS J195316.40+471646.1, is a G5V main--sequence star with an apparent visual magnitude of m$_V$ = 15.69. Spectroscopic analyses indicate an effective temperature of 5746 K and a metallicity of 1.72 dex (\citet{bib27}). \citet{bib25} estimated the stellar age to be approximately 6.3 Gyr and the stellar mass to be 1.125 $M_{\odot}$.

{\sl Kepler--223} hosts a compact planetary system composed of four known sub--Neptune planets, whose physical and orbital properties are summarized in Table 1. The innermost planet, {\sl Kepler--223}b, has a mass of 6.5 M$_{\oplus}$ and a radius of 2.9 R$_{\oplus}$. It orbits the host star with a period of 7.38 days at a semi--major axis of 0.077 au.

The second planet, {\sl Kepler--223}c, has a mass of 6.1 M$_{\oplus}$ and a radius of 3.5 R$_{\oplus}$. Its orbital period is 9.84 days, corresponding to a semi--major axis of 0.093 au.
	
The third planet, {\sl Kepler--223}d, is also a sub--Neptune, with a mass of 8.1 M$_{\oplus}$ and a radius of 5.3 R$_{\oplus}$. It orbits the star every 14.79 days at a distance of 0.123 au.
	
The outermost planet in the system, {\sl Kepler--223}e, has a mass of 4.9 M$_{\oplus}$ and a radius of 4.5 R$_{\oplus}$. It completes one orbit every 19.73 days and is located at a semi--major axis of 0.149 au.

\begin{table*}
\centering
\caption{\label{tab:initial} Initial parameters of the system. We present here the values derived from the gas--driven migration simulations provided by Fabrycky/Siegel, compared with the best--fit values from \citet{bib25}. The initial conditions are given at $T_{epoch}=800.0$ (BJD-2,454,900).}
\setlength{\tabcolsep}{4pt} 
\footnotesize 
\begin{tabular}{@{}lcccccccccccc@{}}
\toprule
 Planet & $P$ & Mass & Radius & $a$ & $e$ & $i$ & $\Omega$ & $\omega_{obs}$ & $M_{obs}$ & $\omega_{gas}$ & $M_{gas}$ \\
  & (days) & ($M_{\oplus}$) & ($R_{\oplus}$) & (au) &  & $(^{\circ})$ & $(^{\circ})$ & $(^{\circ})$ & $(^{\circ})$ & $(^{\circ})$ & $(^{\circ})$ \\
 \midrule
 {\sl Kepler--223}b & 7.3846 & 6.5167 & 2.96 & 0.07719 & 0.0614 & 91.10 & 0.0 & 37.60 & 344.3 & 186.428 & 241.676 \\
 {\sl Kepler--223}c & 9.8456 & 6.0999 & 3.49 & 0.09350 & 0.1124 & 91.08 & 0.0 & 86.06 & 359.6 & 335.514 & 207.423 \\
 {\sl Kepler--223}d & 14.7889 & 8.1239 & 5.29 & 0.12263 & 0.0266 & 91.96 & 0.0 & 58.81 & 274.7 & 100.436 & 13.716 \\
 {\sl Kepler--223}e & 19.7257 & 4.9159 & 4.55 & 0.14860 & 0.0608 & 91.80 & 0.0 & 76.16 & 55.9 & 293.939 & 172.869 \\
\bottomrule
\end{tabular}
\end{table*}
\
\section{Dynamical Analisys}

In this section we describe the numerical methods used to model the Kepler--223 system, with particular emphasis on its long--term evolution and dynamical stability. Our analysis is based on direct N--body integrations of the full equations of motion and is divided into two numerical experiments.

All simulations were performed using the MERCURY N--body package \citet{bib4}, treating all bodies as point masses. We employed the hybrid symplectic integrator throughout, using an initial timestep of 0.03 days (or smaller when required). This timestep corresponds to approximately 1/250 of the orbital period of the innermost planet, {\sl Kepler--223}b (P = 7.38 days), ensuring sufficient accuracy for long--term integrations. When close encounters occur, the hybrid scheme automatically switches to the adaptive Bulirsch--Stoer algorithm, adjusting the timestep to maintain the desired accuracy.

\subsection{Experiment 1: Long--term stability and resonant structure}

In the first numerical experiment, we adopted the orbital solution presented by \citet{bib25} in Extended Data Table 3 (best--fit $C_3$ initial conditions), for the reasons discussed in Section 2. These initial conditions were integrated for 10 Myr, corresponding to approximately $2.0 \times 10 ^8$ orbital periods of the outermost planet, comparable to the integration timescale used by \citet{bib25} to validate their $C_3$ solutions. Over this interval, the system remains stable.

The mean anomalies corresponding to the observational solution were computed following a consistent geometric convention (see Fig.~\ref{fig1}). We adopted the standard coordinate system in which the observer lies along the $z$--axis. A transit occurs when the planet crosses the line of sight, satisfying the condition

\begin{equation*}
	\omega + f = 90^{\circ}
\end{equation*}
where $f$ is the true anomaly. Using transit times closest to our simulation epoch ($T_{epoch} = 800$) to minimize the propagation of TTV uncertainties, we first computed the true anomaly at transit,

\begin{equation*}
	f_{tran} = 90^{\circ} - \omega_{obs}
\end{equation*}
converted it to the corresponding mean anomaly via Kepler’s equation, and then propagated the mean anomaly to the integration epoch using the mean motion. This procedure yields

\begin{equation*}
	M_{obs} \approx 344.3^{\circ}, 359.6^{\circ}, 274.7^{\circ}, 55.8^{\circ}
\end{equation*}
for planets b, c, d, and e, respectively.

However, for the actual N--body integrations presented in this work, we rely exclusively on the gas--driven migration initial conditions ($M_{gas}$ in Table 1) kindly provided by D. Fabrycky. These initial conditions naturally encode the correct resonant phases and yield a configuration located deep within the stable resonant islands.

To verify that the system is indeed locked in resonance, we computed the relevant resonant angles. First, the mean longitudes are defined as

\begin{equation}
\lambda_1=M_1+\omega_1+\Omega_1
\end{equation}
\begin{equation}
\lambda_2=M_2+\omega_2+\Omega_2
\end{equation}
\begin{equation}
\lambda_3=M_3+\omega_3+\Omega_3
\end{equation}

We then evaluated the two--body resonant angles between adjacent planets:
\begin{equation}
\theta_{1 2}= 3 \lambda_1-4  \lambda_2 + \varpi_2
\label{eq01}
\end{equation}

\begin{equation}
\theta_{2 1}= 3 \lambda_1-4  \lambda_2 + \varpi_1
\end{equation}

\begin{equation}
\theta_{2 3}= 2 \lambda_2-3  \lambda_3 + \varpi_3
\end{equation}

\begin{equation}
\theta_{3 2}= 2 \lambda_2-3  \lambda_3 + \varpi_2
\end{equation}

\begin{equation}
\theta_{3 4}= 3 \lambda_3-4  \lambda_4 + \varpi_4
\end{equation}

\begin{equation}
\theta_{4 3}= 3 \lambda_3-4  \lambda_4 + \varpi_3
\end{equation}

as well as the three--body resonant angles,
\begin{equation}
\phi_{1 2 3}= - \lambda_1+2  \lambda_2 - \lambda_3
\end{equation}

\begin{equation}
\phi_{2 3 4}= \lambda_2-3  \lambda_3 + 2 \lambda_4
\end{equation}
and the four--body Laplace--type angle,

\begin{equation}
\phi_{1 2 3 4}=  3 \lambda_1-4\lambda_2-3  \lambda_3 + 4 \lambda_4
\label{eq08}
\end{equation}

Additionally, we explored nine nearby locations in the (a, e) space of {\sl Kepler--223}b, within the observational uncertainties of the $C_3$ solution, to examine how the resonant angles behave in the immediate neighborhood of the nominal configuration.

\subsection{Experiment 2: Stability maps in the $(a, e)$ plane}

In the second numerical experiment, we used the long--term stable configuration identified in Experiment 1 as a reference solution and constructed two--dimensional dynamical maps in the $(a, e)$ plane for each planet. These maps are designed to reveal the structure of the mean--motion resonances and the surrounding stability islands in phase space.

Following previous studies of resonant systems (e.g., Kepler--60: \citet{bib9}; Kepler--36: \citet{bib38}), we selected one planet at a time and fixed the orbital parameters of the remaining three. For the chosen planet, we constructed a grid of $132$ $\times$ $100$ initial conditions spanning a range of eccentricities from e = 0 to 0.9 and a semi–major axis interval of

\begin{equation*}
	a \pm 2 R_H
\end{equation*}
where $R_H$ is the planet's Hill radius

Although the classical instability region for non--resonant interacting planets is often estimated to extend to approximately three mutual Hill radii \citealp{bib48}, we deliberately restricted our search to $\pm 2R_H$. This choice allows us to resolve the fine structure of the resonant stability islands, which are expected to be confined to this region, rather than sampling the broader chaotic domain where instability is already guaranteed.

Each grid point was integrated for 100,000 years, corresponding to $1.85 \times 10^6$ orbits of the outermost planet. This integration length matches that used by \citet{bib9} for Kepler--60 and is sufficient to detect short--term chaotic behavior associated with MMR--driven instability \citep{bib8}. If an integration resulted in a collision (planet--planet or planet--star) or an ejection, the corresponding $(a, e)$ pair was classified as unstable, and the event time was recorded from MERCURY’s output.

The resulting $(a, e)$ phase--space maps provide a powerful diagnostic of the dynamical environment surrounding each planet, offering insight into the depth of resonance, the extent of phase--space protection, and the proximity of the system to chaotic boundaries.

For the second experiment, we used the initial conditions that we found to be long--term stable (experiment 1) in our first experiment and searched for unstable areas on the $(a, e)$ plane ($a$ being the initial planetary semi--major axis and $e$ being the initial eccentricity) for each planet, that is 2--dim dynamical maps in the neighborhood of the best fitting solution with $C_3$. This experiment is performed to show the MMR structure of the phase--space around the solution (as it was done for Kepler--60 in \citet{bib9}, Kepler--36 in \citet{bib38}). We choose one of the four planets and fix all the orbital parameters of the other three. Then we constructed a grid of $132$ $\times$ $100$ models corresponding to the same number of $(a, e)$ points, with $e$ ranging from $0$ to $0.9$, For the semi--major axis range, we selected an interval of $a \pm 2 R_H$ (where $R_H$ is the planet's Hill radius). Although the theoretical instability zone for non--resonant interacting planets is typically estimated to extend up to $\sim 3$ mutual Hill radii (e.g., \citealp{bib48}), we restricted our high--resolution search to $\pm 2 R_H$. This choice allows us to focus specifically on resolving the fine structure of the resonant stability islands, which are expected to be tightly confined within this region, rather than exploring the broader chaotic sea where instability is already guaranteed.
We integrated numerically each initial condition of the grid for 100,000 years, corresponding to  $1.85 \times 10^6$ orbits of the outermost planet, this number of orbits is the same that \citet{bib9} used for Kepler--60 and is sufficient to detect short--term chaotic motions for the MMR instability time--scale \citep{bib8}. 
Then in case a given initial position leads to an unstable planetary orbit (either collision between planets, ejections or hit the central star), the specific $(a, e)$ pair is classified as unstable and moved on to the next pair of values,  the collision time corresponding to that unstable is saved in the info.out file in Mercury.
	
The Mean Anomalies are important to find the different resonant angles and how they librate. The phase--space is an excellent tool to understand the neighbourhood of the resonance in which each planet is (not only the objects that are close to the  exact MMR), the phase space a--e gives very important information. As an example of this, we cite \citet{bib38}, where they study Kepler--36. First, they explored if the system configuration could be obtained by interactions with the protoplanetary disc.  In their section 5.0 named "Stability", subsection 5.2 "Overview of the parameter space", they state "all angles are chosen from a uniform distribution."

\begin{figure}[h]
    \centering
    \begin{tikzpicture}[scale=2.5, >=Latex]
        \def\e{0.5} 
        \def\w{35}  
        \def\f{55}  
        
        \draw[dashed, gray] (0,0) -- (2.5,0) node[right, black] {Reference Line (Node)};
        
        \draw[->, thick, red] (0,0) -- (\w:2.0) node[above right] {Laplace Vector ($\mathbf{e}$)};
        \node[red, rotate=\w] at (\w:1.2) {\footnotesize Periapsis};

        \draw[->, thick, blue] (0,0) -- (\w+\f:2.2) node[above] {To Observer ($Z$-axis)};
        \fill[blue] (\w+\f:1.5) circle (0.08) node[right=4pt] {Planet};
        \node[blue] at (\w+\f:1.8) {\footnotesize (Line of Sight)};
        
        \draw[->] (0.6,0) arc (0:\w:0.6) node[midway, right] {$\omega$};
        
        \draw[->] (\w:0.7) arc (\w:\w+\f:0.7) node[midway, above] {$f$};
        
        \node[align=left, anchor=north west] at (-1.5, -0.5) {
            \textbf{Transit Condition:}\\
            Observer on $Z$-axis\\
            $\omega + f = 90^{\circ}$
        };
        
        \fill[orange] (0,0) circle (0.12) node[below=5pt, black] {Star};
        
        \draw[thick, gray, dashed] (\w-30:1.5) arc (\w-30:\w+\f+30:1.5);
        
    \end{tikzpicture}
    \caption{Sketch of the orbital plane defining the coordinate system used in this work (following \citet{bib45} and using the notation of \citet{bib28}). The observer is located along the $z$-axis (Line of Sight). The mean anomaly is measured from the \textbf{Laplace vector} (eccentricity vector $\mathbf{e}$), which points to the periapsis. }
    \label{fig1}
\end{figure}

\begin{figure}
\includegraphics[width=\columnwidth]{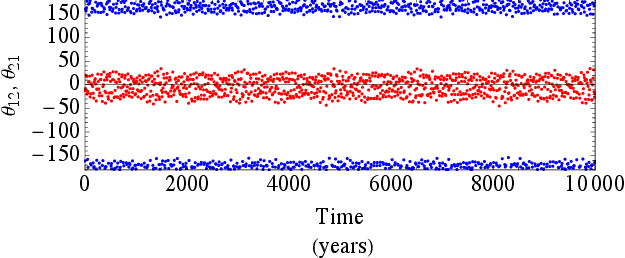}
\caption{Here we show the two--body resonant angle $\theta_{1 2}$, as defined in Eq. 6 (shown in red), and  $\theta_{2 1}$ as defined in  Eq. 7 (shown in blue). Both angles are librating, $\theta_{1 2}$ around $0^{\circ}$ and $\theta_{2 1}$ around 180$^{\circ}$, see text for full explanation. }
\label{fig2}
\end{figure}

\begin{figure}
\includegraphics[width=\columnwidth]{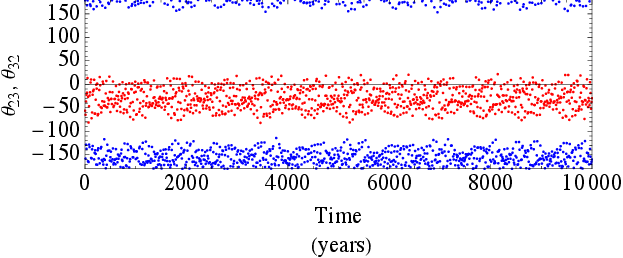}
\caption{Here we show the two--body resonant angle $\theta_{2 3}$, as defined in Eq. 8 (shown in red), and  $\theta_{3 2}$ as defined in  Eq. 9 (shown in blue). Both angles are librating, $\theta_{2 3}$ around $0^{\circ}$ and $\theta_{3 2}$ around 180$^{\circ}$, see text for full explanation. }
\label{fig3}
\end{figure}

\begin{figure}
\includegraphics[width=\columnwidth]{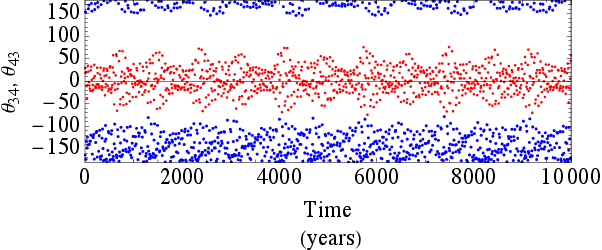}
\caption{Here we show the two--body resonant angle $\theta_{3 4}$, as defined in Eq. 10 (shown in red), and  $\theta_{4 3}$ as defined in  Eq. 11 (shown in blue). Both angles are librating, $\theta_{3 4}$ around $0^{\circ}$ and $\theta_{4 3}$ around 180$^{\circ}$, see text for full explanation.}
\label{fig4}
\end{figure}

\begin{figure}
\includegraphics[width=\columnwidth]{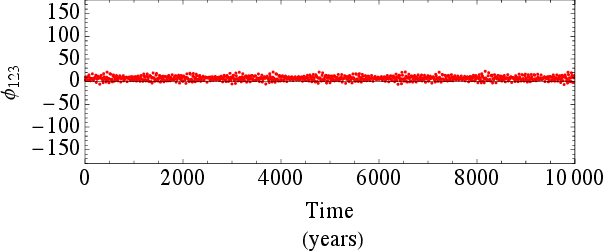}
\caption{Here we show the three--body resonant angle $\phi_{1 2 3}$, as defined in Eq. 12 (shown in red), as can be seen the resonant angle librates around zero in this plot, and also have short periods of circulation, see text for details.}
\label{fig5}
\end{figure}

\begin{figure}
\includegraphics[width=\columnwidth]{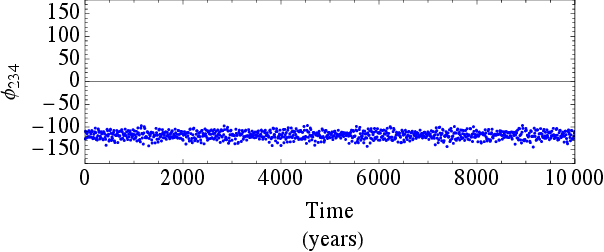}
\caption{Here we show the three--body resonant angle $\phi_{2 3 4}$, as defined in Eq. 13 (shown in blue), as can be seen the resonant angle librates around $180^{\circ}$ in this plot, Is not as well defined as in Figure 4, but still is clear the system in at resonance.}
\label{fig6}
\end{figure}

\begin{figure}
\includegraphics[width=\columnwidth]{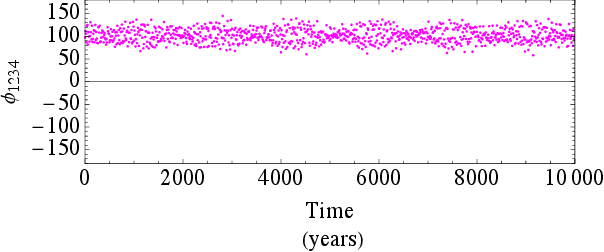}
\caption{Here we show the four--body resonant angle $\phi_{1 2 3 4}$, as defined in Eq. 14 (shown in purple), as can be seen the resonant angle does not seem to librate but rather circulate.}
\label{fig7}
\end{figure}


\begin{figure}
\includegraphics[width=\columnwidth]{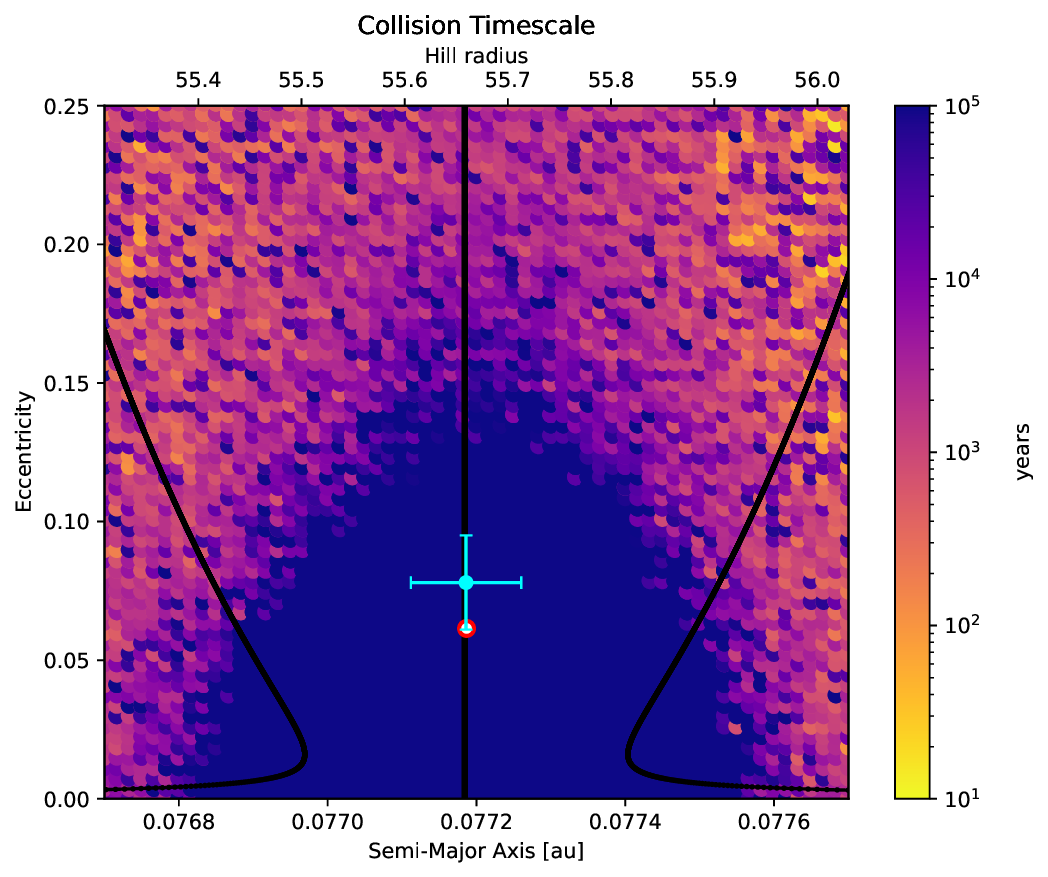}
\caption{Results of the dynamical environment of {\sl {\sl Kepler--223}b}, represented here as a red circle, the colour bar represents the collision time as a measurement of chaos and unstability in the system, blue representing larger collision times and yellow representing short collision times, the planet is within a stability island, we also represent the error bars in light blue, see text for details.}
\label{fig8}
\end{figure}


\begin{figure}
\includegraphics[width=\columnwidth]{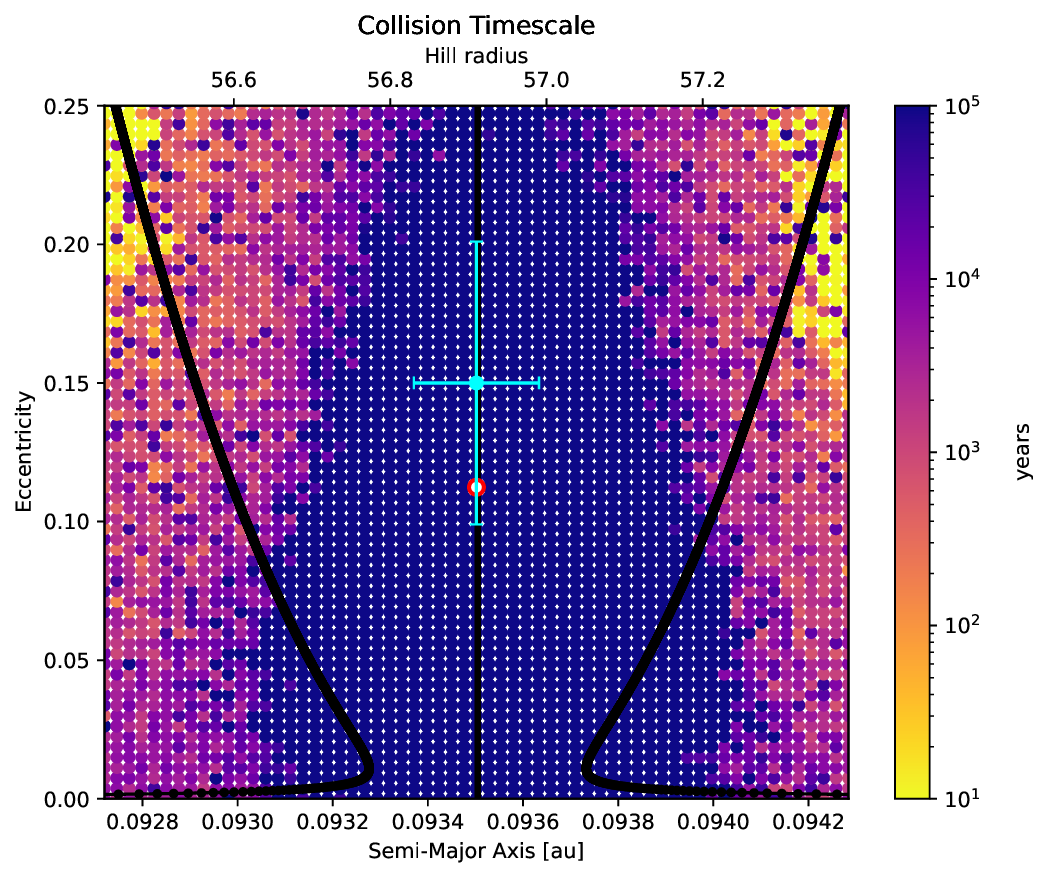}
\caption{Results of the dynamical environment of {\sl {\sl Kepler--223}c}, represented here as a red circle, the colour bar represents the collision time as a measurement of chaos and unstability in the system, blue representing larger collision times and yellow representing short collision times, the planet is within a stability island, we also represent the error bars in light blue, see text for details.}
\label{fig9}
\end{figure}

\begin{figure}
\includegraphics[width=\columnwidth]{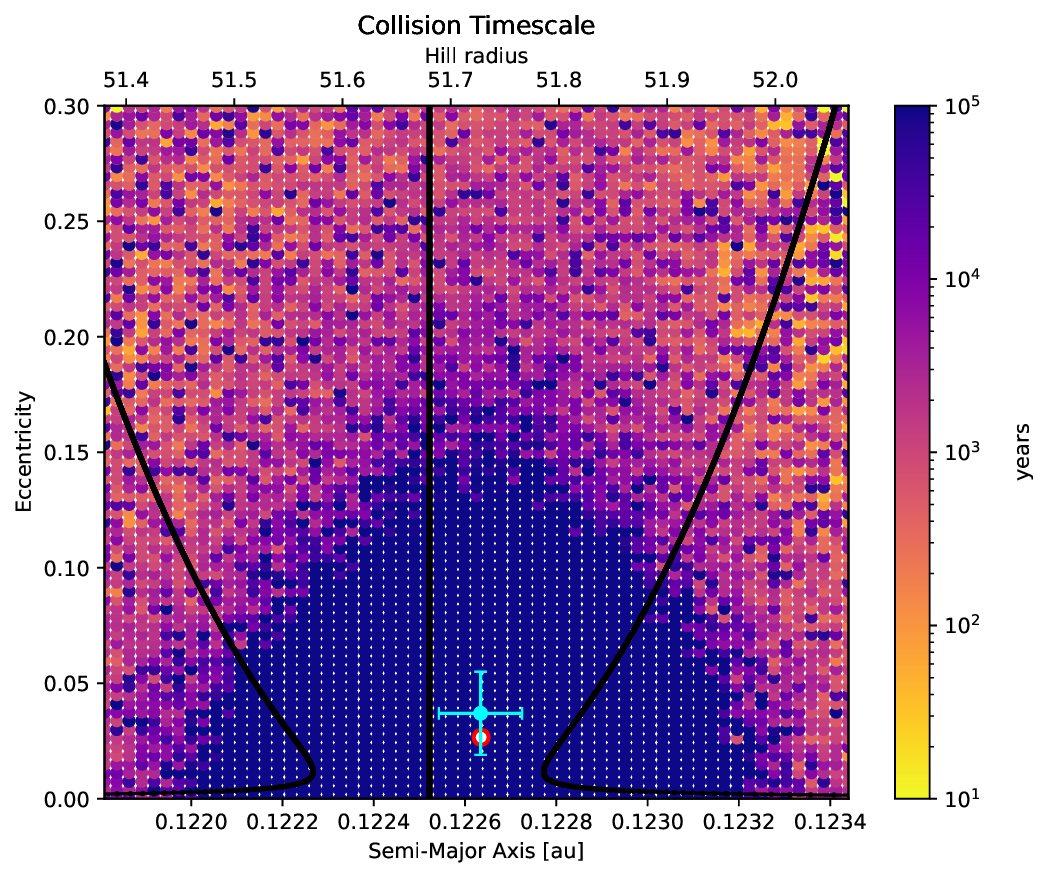}
\caption{Results of the dynamical environment of {\sl {\sl Kepler--223}d}, represented here as a red circle, the colour bar represents the collision time as a measurement of chaos and unstability in the system, blue representing larger collision times and yellow representing short collision times, the planet is within a stability island, we also represent the error bars in light blue, see text for details.}
\label{fig10}
\end{figure}

\begin{figure}
\includegraphics[width=\columnwidth]{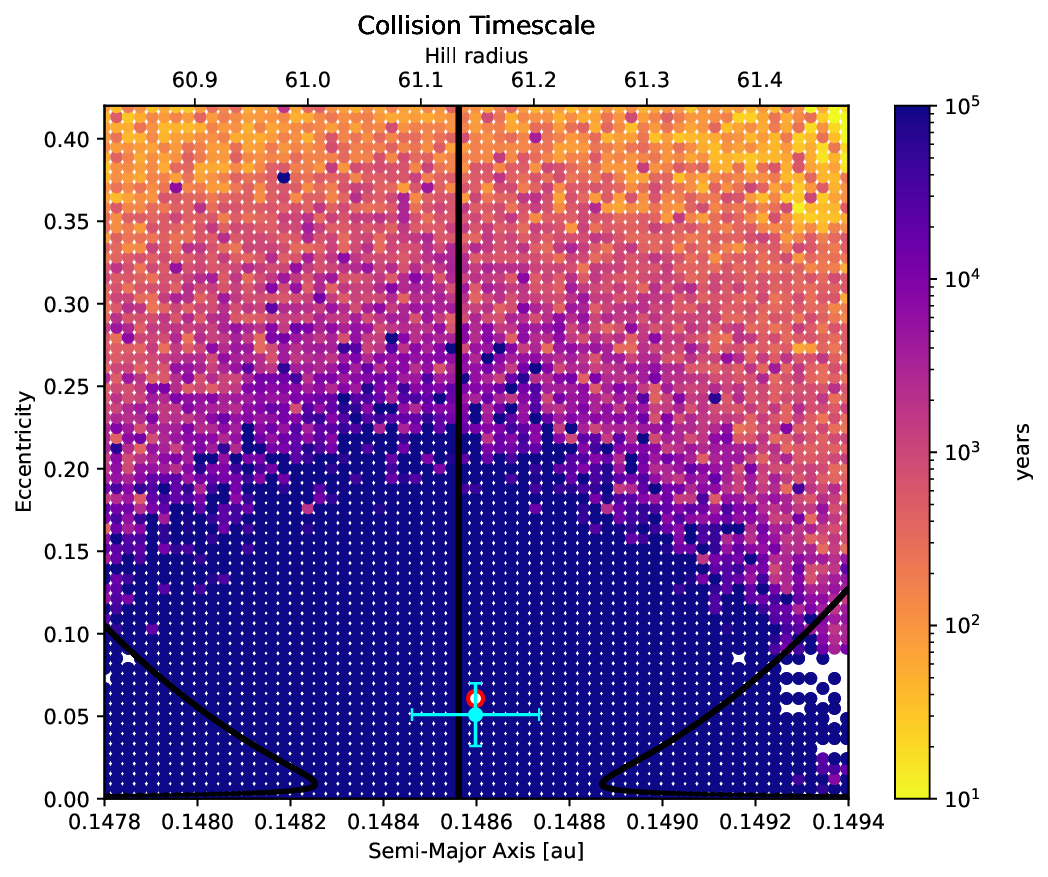}
\caption{Results of the dynamical environment of {\sl {\sl Kepler--223}e}, represented here as a red circle, the colour bar represents the collision time as a measurement of chaos and unstability in the system, blue representing larger collision times and yellow representing short collision times, the planet is within a stability island, we also represent the error bars in light blue, see text for details.}
\label{fig11}
\end{figure}

\section{Results}
\subsection{Long--term results} 

We first present the results of our long--term numerical integrations. As described in the previous sections, we integrated forward in time the orbital solution reported by \citet{bib25}. Specifically, we adopted the best--fit initial conditions listed in the bottom rows of Extended Data Table 3, corresponding to the $C_3$ solution obtained through the DEMCMC analysis under dynamical stability constraints.

Our simulations show that this configuration remains dynamically stable over an integration timespan of 10 Myr, corresponding to approximately $2.0 \times 10 ^8$ orbital periods of the outermost planet.

\subsection{Resonant Angles} 

Using the same initial conditions adopted in the previous subsection (Table 1), we performed additional numerical integrations over $10^5$ years, focusing on the dynamical behavior of the resonant angles associated with the two--body, three--body, and four--¸body resonances defined in Section 4.

We first examine the two--body resonant angles $\theta_{12}$,  $\theta_{23}$,  $\theta_{32}$, and  $\theta_{43}$, shown in Figs. 2--4. In each case, one resonant angle (shown in red) librates about $0^{\circ}$, while the corresponding angle (shown in blue) librates about $180^{\circ}$. This behavior is consistently observed for all adjacent planet pairs, indicating that the entire system is locked in two--body mean--motion resonances.

We next examine the three--body resonant angles $\phi_{1 2 3}$, and $\phi_{2 3 4}$, shown in Figs. \ref{fig5}--\ref{fig6}, respectively. Figure \ref{fig5} illustrates the evolution of $\phi_{1 2 3}$, which initially librates about $0^{\circ}$, subsequently shifts toward $180^{\circ}$, and later returns to libration around $0^{\circ}$. Figure \ref{fig6} displays the behavior of $\phi_{2 3 4}$, which predominantly librates around $180^{\circ}$. Although the libration in this case is less regular and exhibits intermittent circulation, the overall behavior clearly indicates that the system remains locked in this three--body resonance.

Finally, we examine the four--body resonant angle $\phi_{1 2 3 4}$, shown in Fig. \ref{fig7}. In contrast to previous studies based on best--fit posterior samples, the gas--driven migration initial conditions used here reveal a clear and robust libration of this angle about $0^{\circ}$ (or $180^{\circ}$, depending on the chosen phase convention). This libration persists over all timescales explored (10,000, 100,000 and 1Myrs), demonstrating that {\sl Kepler--223} resides in a deep, long--lived four--body resonant chain.

As mentioned in the previous subsection, we explored nine representative points in the $(a,e)$ phase space around the $C_3$ solution of {\sl Kepler--223}b, within the observational uncertainties, in order to assess the sensitivity of the resonant angles to variations in these two parameters. The results of this analysis are presented in Appendix~\ref{Append}.

\subsection{Stability Maps}

In this subsection, we present the results of our second numerical experiment, namely the construction of stability maps for the known planets in the {\sl Kepler--223} system. To define and quantify orbital stability, we follow the framework introduced by \citet{bib38}.

According to \citet{bib38}, planetary stability can be characterized using several complementary metrics, each associated with a different timescale over which instability may manifest. In particular, four characteristic timescales are commonly considered:

i) Lyapunov timescale: This measures the time required for two initially nearby trajectories in phase space to diverge exponentially. A short Lyapunov time is indicative of chaotic dynamics, which often—but not necessarily—leads to macroscopic instability.

ii) Collision timescale: This corresponds to the time elapsed until two planets undergo a physical collision. In our simulations, this quantity is computed automatically by the Mercury integrator and recorded in the info.out output file.

iii) Ejection timescale: This measures the time until a planet becomes gravitationally unbound from the system. In our integrations, as also reported by \citet{bib38}, instability in compact resonant chains typically manifests through collisions rather than planetary ejections.

iv) Lagrange timescale: This refers to the duration over which the orbital elements remain bounded. Although formally defined as confinement within a specific region of phase space, in practical numerical studies this timescale is often quantified by monitoring significant variations in the semi--major axes of the planets.

In their study of the {\sl Kepler--36} system, \citet{bib38} compared several stability diagnostics derived from N--body simulations. In their Figure 8, they present three stability maps based on different metrics: the Lyapunov timescale (left panel), the Lagrange timescale (middle panel), and the collision timescale (right panel). The close qualitative agreement among these maps demonstrates that all three diagnostics convey essentially the same dynamical information and are therefore equally valid measures of stability.

In the present work, we adopt the collision timescale as our primary stability metric. For compact planetary systems locked in deep mean--motion resonances, such as {\sl Kepler--223}, dynamical instability almost invariably manifests through orbit crossings followed by physical collisions rather than planetary ejections. Recent studies, including \citet{bib40}, have similarly emphasized planet--planet collisions as the most robust and unambiguous criterion for instability in tightly packed resonant architectures.

A critical methodological distinction of this work is our employment of the hybrid symplectic/Bulirsch--Stoer integrator within the MERCURY package. Unlike purely symplectic schemes used in prior large--scale surveys, the hybrid algorithm adaptively reduces the timestep (initially set to 0.03 days, $\sim P_b/250$) during close encounters. This adaptive switching ensures that the instability events recorded in our maps are physically robust collisions rather than numerical artifacts arising from fixed--step errors. Consequently, our collision timescales represent a high--fidelity metric of the system's true dynamical limits.

This hybrid scheme operates with a fast symplectic algorithm when planets remain well separated (in units of mutual Hill radii), and automatically switches to the adaptive Bulirsch–Stoer integrator during close encounters. The Bulirsch–Stoer method, with its variable time step and user--controlled accuracy, is particularly well suited for resolving close approaches and collision events, where fixed--step symplectic methods may lose precision.

As a consequence, close encounters in our simulations are modeled with higher numerical fidelity, and the resulting collision times are physically robust rather than numerical artifacts. This choice is especially important for compact, resonant systems such as {\sl Kepler--223}, where instability typically manifests through orbit crossings and collisions, and where accurate treatment of short--timescale dynamics is essential.

We identify both dynamically stable regions (shown in blue tones) and unstable regions (yellow tones) in the $(a,e)$ phase space for each planet, using the collision timescale as our primary diagnostic of chaos and instability. The resulting stability maps are presented in Figures~\ref{fig8} to~\ref{fig11}. In each panel, the color bar on the right indicates the collision timescale encoded in color: yellow regions correspond to very short collision times, whereas blue regions indicate long collision times and long--term stability.

The location of the solution reported by  \citet{bib25} in their Extended Data Table 3 (best--fit to the data initial conditions under the $C_3$ constraints; bottom values in the table) is marked by a red circle in each map. Each grid point in the $(a,e)$ plane was integrated for $10^5$ yr, corresponding to approximately $1.85 \times 10^6$ orbits of the outermost planet.

We overlay the observational constraints from \citet{bib25} (light blue error bars) onto our dynamical maps to contextualize the system's architecture. It is important to distinguish between the \textit{observational snapshot}—which constrains the osculating elements at the current epoch—and the \textit{dynamical environment} revealed by our maps. By exploring the phase space within $\pm 2$ Hill radii, we demonstrate that the Kepler--223 planets do not merely exist within the observational error bars by chance; rather, they inhabit specific, stability islands carved out by the system's resonance history.

In this work, however, we deliberately explore the semi--major axis and eccentricity beyond the formal observational uncertainties, sampling the surrounding phase space on scales of a fraction of each planet’s Hill radius. This approach, commonly adopted in dynamical studies (e.g. \citet{bib9},  \citet{bib24}, \citet{bib38}), allows us to characterize the local dynamical environment in which each planet resides and to assess the robustness of the resonant configuration against small perturbations.

Finally, the nominal locations of the relevant mean--motion resonances, computed following \citet{bib28}, Eq. (8.22), are indicated by vertical black lines in Figures~\ref{fig8} to~\ref{fig11}.

Figure~\ref{fig8} presents the stability map for {\sl Kepler--223}b. The nominal solution (red circle) is embedded within a well--defined region of long--term stability, shown in blue, which forms a clear stability island in the $(a,e)$ plane. Although a small number of unstable solutions (yellow points) appear both near the boundary and sporadically within the island, the system is unambiguously located deep inside the stable region. The nominal location of the relevant mean--motion resonance (MMR) is indicated by a vertical black line, and {\sl Kepler--223}b lies very close to this resonant location.

Figure~\ref{fig8} shows the stability map for {\sl Kepler--223}c. In this case, the stability island is particularly well defined, with almost no unstable (yellow) solutions inside the island. The best--fit solution (red circle) lies well within the interior of the stable region, far from its boundaries, indicating a robust and resilient configuration. As for {\sl Kepler--223}b, the planet’s nominal orbit is located very close to the MMR, as indicated by the vertical black line.

The stability map for {\sl Kepler--223}d is shown in Figure~\ref{fig10}. Here, the stability island is less sharply defined than for the two inner planets and exhibits a more complex boundary structure. Several unstable solutions appear within the island; nevertheless, the nominal solution (red circle) is clearly surrounded by stable (blue) orbits, confirming that {\sl Kepler--223}d resides within a region of long--term stability. In this case, the MMR location (black vertical line) lies slightly interior to the planet’s nominal semi--major axis.

Finally, Figure~\ref{fig11} presents the stability map for {\sl Kepler--223}e. The nominal solution again lies well within a stable island, despite the presence of a small number of unstable solutions embedded within it. As for {\sl Kepler--223}d, the boundary of the stability island is intricate, reflecting the complex resonant dynamics of the outer planet. The location of the corresponding MMR is found interior to the nominal orbit of {\sl Kepler--223}e.

What Figures~\ref{fig8} to~\ref{fig11} have all in common is that reveal that the best-fit solution is not merely stable, but deeply embedded in the resonance. The stability island fully accommodates the observational error bars, isolating the system from chaotic zones.

To justify our choice of integration time for the stability maps, we performed an additional set of simulations extending the integrations to $10^6$ yr for the same grid of initial conditions shown in Figures~\ref{fig8}. we found only minimal differences compared to the $10^5$ yr integrations, confirming that our adopted timescale is sufficient to robustly identify the stable and unstable regions of phase space.

\subsection{Resonance Width}

Following Murray \& Dermott (1999), and as further elaborated in \citet{bib38}, if only the resonant terms are retained, the perturbing force exerted by a planet on a test particle can be expressed as a series expansion in the orbital elements. Using the notation adopted in this work, the general form of the resonant argument (angle) between body $k$ and planet $l$, denoted as $\theta_{k l}$, can be written as

\begin{equation}
	\theta_{k l}= j_1 \lambda_k + j_2  \lambda_l + j_3 \omega_k +  j_4 \omega_l+ j_5 \Omega_k + j_6 \Omega_l
\end{equation}

The specific resonant angle is determined by the values of the integer coefficients $j_1$--$j_6$. These coefficients are used to evaluate Eq. (24) of \citet{bib38} for each of the resonant arguments appearing in Eqs. \ref{eq01}--\ref{eq08}.

We compute the libration widths for two--body resonances between adjacent planets. Since our analysis is restricted to first--order resonances, we employ the corresponding two--body resonant angles to calculate the libration width $\delta a_{max}$ as a function of eccentricity, following Eq. (26) of \citet{bib38}. For {\sl Kepler--223}b, we first evaluated Eq. (24) of \citet{bib38} and subsequently derived $\delta a_{max}$ using Eq. (26) of the same work.

The results are shown in Figures~\ref{fig8} to~\ref{fig11}, where the libration widths are $\delta a_{max}$ represented by black curves on either side of the vertical black line indicating the nominal resonance location. As can be seen, the curves provide a good estimate of the mean width of the associated stability island in each case. A more detailed discussion of these results is presented in the following section.
\\
\\

\section{Discussion}
\label{Sec:Discussion}

We investigated the dynamical stability of the four known planets in the {\sl Kepler--223} system through direct numerical integrations of the full equations of motion. Our analysis was divided into two complementary experiments: (i) a long--term integration of the nominal orbital solution and (ii) a systematic exploration of the phase space via stability maps.

In the first experiment, we integrated the updated initial conditions provided by D. Fabrycky—derived from \citet{bib25} and subsequent refinements—for a total duration of 10 Myr. We confirm that this solution remains stable for at least $2.0 \times 10^8$ orbital periods of the outermost planet. One of the key results of this study concerns the behavior of the resonant angles. In contrast to previous works, in which some higher--order resonant angles were found to circulate, our simulations show that all two--body and three--body resonant angles, as well as the four--body resonant angle $\phi_{1234}$, exhibit persistent libration over the entire integration timespan.

This behavior demonstrates that the {\sl Kepler--223} system is not merely located near a resonant configuration, but is instead trapped in a deep and robust four--body resonant chain. Such a configuration acts as a dynamical protection mechanism, preventing close planetary encounters and ensuring long--term stability.

In our second experiment, we constructed stability maps in the $(a,e)$ plane (Figures~\ref{fig8} to~\ref{fig11}). These maps show that each planet occupies a well--defined stability island associated with its corresponding mean--motion resonance (MMR). Using the updated initial conditions, we find that the stable regions (shown in blue) are located closer to the exact nominal resonance positions (vertical black lines) than in previous estimates. Moreover, the stability islands are not strictly confined to low eccentricities; instead, stable solutions extend to significantly higher eccentricity values, indicating that the resonant phase--protection mechanism operates over a broader region of phase space.

Our N--body results are in excellent agreement with the analytical first--order resonance widths predicted by \citet{bib46}, shown as black curves in the figures. The sharp transition from regular to chaotic motion at the boundaries of these islands is consistent with the resonance overlap criterion described by \citet{bib47}. For {\sl Kepler--223}b, c, and d, the numerically derived stability islands are well contained within the corresponding analytical boundaries. In contrast, for the outermost planet, {\sl Kepler--223}e (Fig. \ref{fig11}), the stable region extends significantly beyond the analytically predicted width. This discrepancy suggests that, for the outer member of the resonant chain, higher--order resonant terms or complex multi--body interactions induced by the inner planets may effectively broaden the stable region beyond standard first--order analytical expectations.

Finally, the long collision timescales measured within these stability islands further confirm the robustness of the phase--protection mechanism, thereby justifying our use of full N--body integrations rather than fast chaos indicators for assessing the long--term stability of the system.

\subsection{Comparative Dynamics}
The dynamical architecture of Kepler--223 offers a distinct counterpoint to other known resonant chains. For instance, while the TRAPPIST-1 system \citep{bib49} hosts a long chain of seven terrestrial planets, Kepler--223 distinguishes itself by the higher mass of its components (sub--Neptunes) and its evolved age ($\sim 6$ Gyr). A more recent analogue is the TOI-178 system, where five outer planets form a resonant Laplace chain \citep{bib51}. However, unlike TOI-178, which exhibits significant density variations suggesting a complex formation history, Kepler--223 appears to be a pristine `fossilized' gravitational clockwork. Its structure has preserved the primordial fingerprints of convergent migration for billions of years without the need for the active tidal dissipation mechanisms that maintain the Galilean moons of Jupiter \citep{bib50}.

\section{Conclusions}
\label{Sec:Conclusions}

In this work, we have carried out a detailed dynamical characterization of the four--planet resonant chain {\sl Kepler--223} using updated initial conditions. Our main conclusions can be summarized as follows:

\begin{enumerate}
	\item Deep Resonant Lock\\
	The system is dynamically stable on long timescales and exhibits robust libration of all resonant arguments. Most importantly, we confirm the persistent libration of the four–body resonant angle $\phi_{1234}$ demonstrating that the planets are trapped in a genuine, deeply locked multi--planet resonant chain.	
	\item Stability Islands\\
		Each planet resides within a well--defined stability island centered near the exact mean--motion resonance locations. Even when accounting for observational uncertainties. These islands are robust and extend to high eccentricities, showing that the system's stability is preserved, indicating that the long--term stability of the system is maintained over a broader region of phase space.
	\item Analytical Agreement and Deviations \\
		For the inner three planets ({\sl Kepler--223}b, c, and d), the location and width of the stability islands are well described by the analytical theory of first--order resonance widths. In contrast, the outermost planet ({\sl Kepler--223}e) exhibits a stability region that is wider than analytically predicted, underscoring the importance of higher–order effects and complex multi--body interactions at the outer ¨edge of the resonant chain.
\end{enumerate}



\newpage
\clearpage
\begin{appendices}



\section{} 
\label{Append}
Here we show the 9 locations in the a--e space for planet {\sl Kepler--223b}, this figure is just a zoom of the area with the error bars that appears in Figure~\ref{fig8}  that are within the error bar area around the initial condition to see the behaviour of the resonant angles around the C3 solution {\sl Kepler--223b}. We do this to see the effect that small changes on a--e parameters have on the three--body resonant angles that are shown in this Appendix.

\begin{figure}[b]
\includegraphics[width=8.1cm]{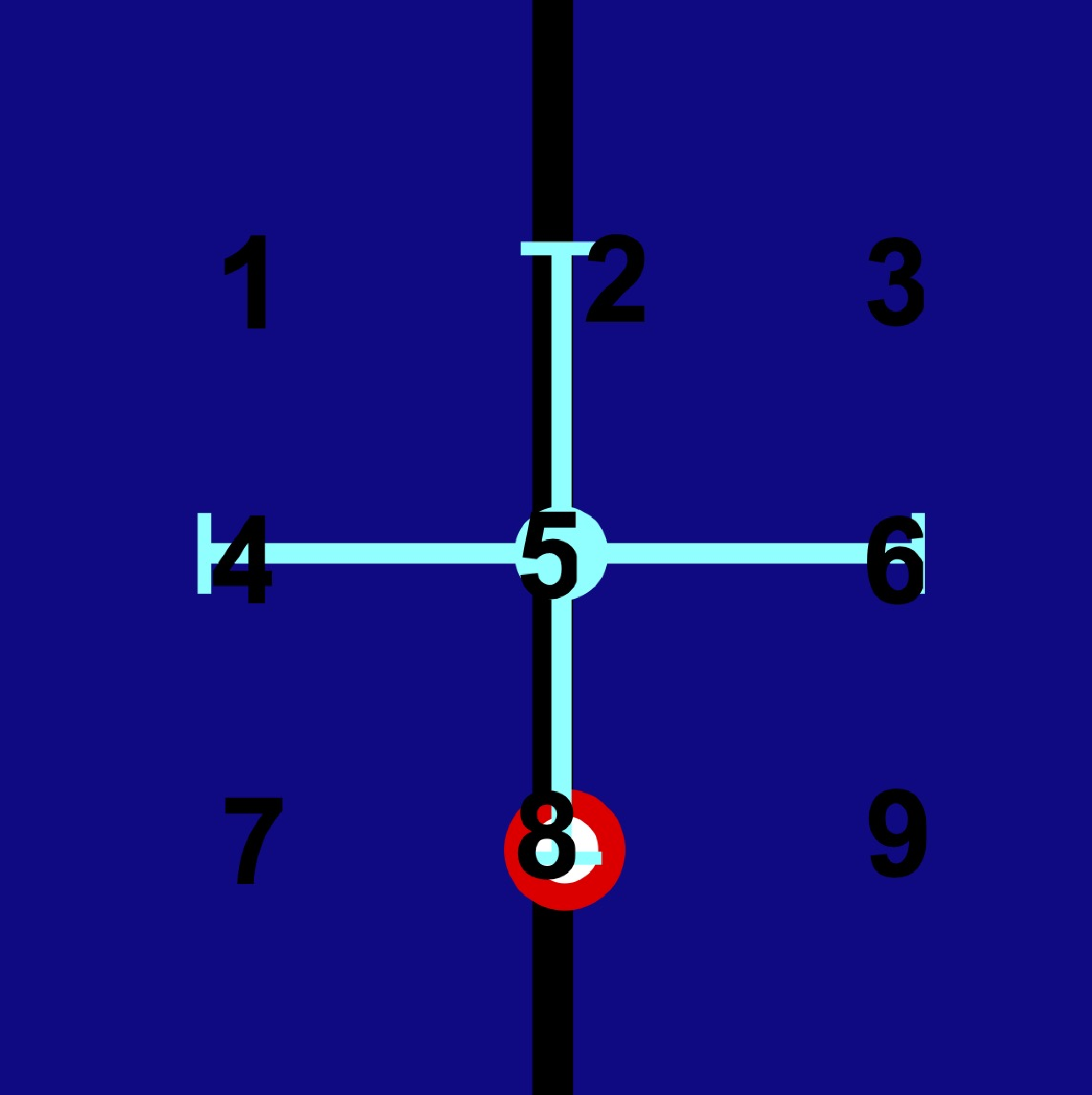}
\caption{Locations in the a--e space of the 9 points around the C3 initial condition for {\sl Kepler--223b} to see the effect that small changes on a--e parameters have on the three--body resonant angles that are shown below in this section, this figure is just a zoom of the area with the error bars that appears in Figure~\ref{fig8}.}
\label{A1}
\end{figure}

\begin{figure}[b]
\begin{multicols}{2}
    \includegraphics[width=6.8cm]{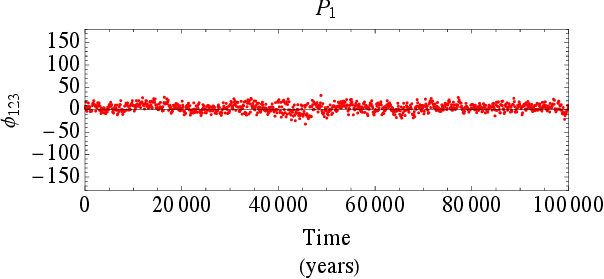}\par 
    \includegraphics[width=6.8cm]{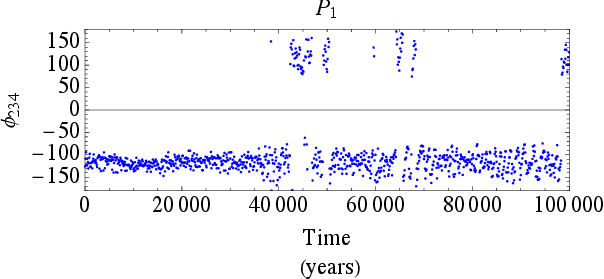}\par
    \end{multicols}
\begin{multicols}{2}
    \includegraphics[width=6.8cm]{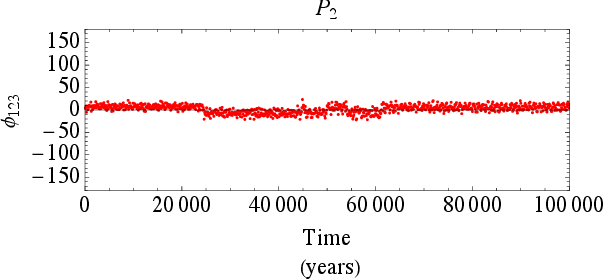}\par
    \includegraphics[width=6.8cm]{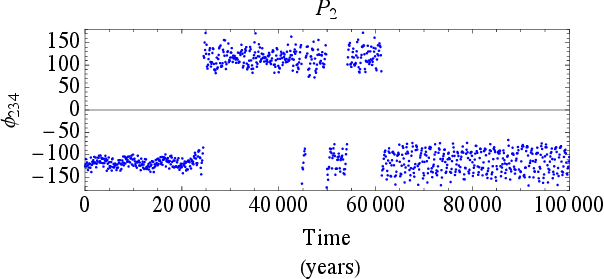}\par
\end{multicols}
\begin{multicols}{2}
    \includegraphics[width=6.8cm]{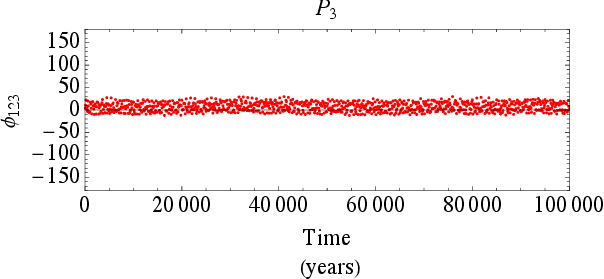}\par
    \includegraphics[width=6.8cm]{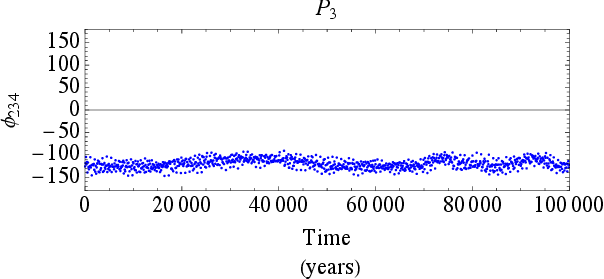}\par
\end{multicols}
\caption{Three--body resonant angles, $\phi_{1 2 3}$ (shown in red) and $\phi_{2 3 4}$ shown in blue,  corresponding to the points 1,2 and 3 that appear in Figure~\ref{A1}. It is clear that small changes in the a--e parameters have changes in these two angles}
\end{figure}

\begin{figure}[b]
\begin{multicols}{2}
    \includegraphics[width=6.8cm]{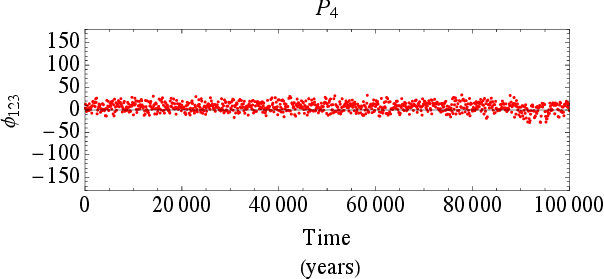}\par 
    \includegraphics[width=6.8cm]{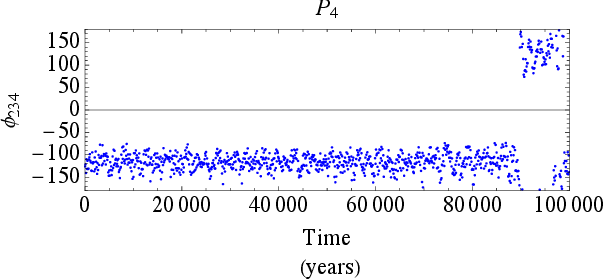}\par
    \end{multicols}
\begin{multicols}{2}
    \includegraphics[width=6.8cm]{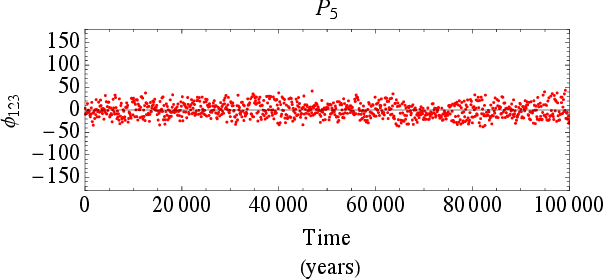}\par
    \includegraphics[width=6.8cm]{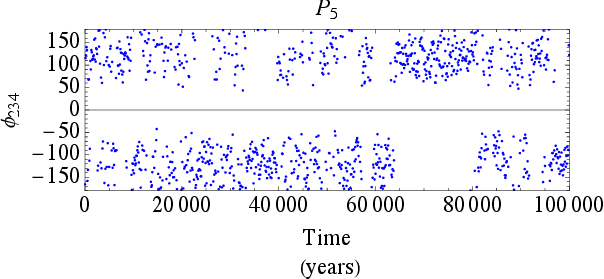}\par
\end{multicols}
\begin{multicols}{2}
    \includegraphics[width=6.8cm]{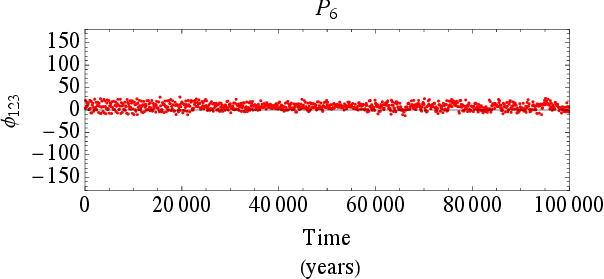}\par
    \includegraphics[width=6.8cm]{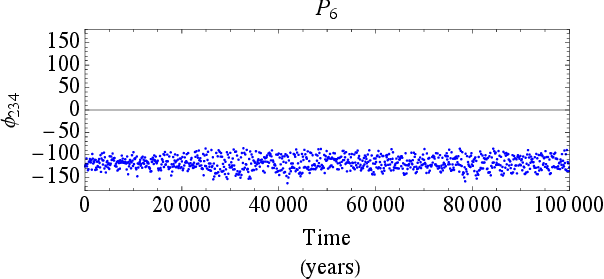}\par
\end{multicols}
\caption{Three--body resonant angles, $\phi_{1 2 3}$ (shown in red) and $\phi_{2 3 4}$ shown in blue,  corresponding to the points 4,5 and 6 that appear in Figure~\ref{A1}. It is clear that small changes in the a--e parameters have changes in these two angles}
\end{figure}

\begin{figure}[b]
\begin{multicols}{2}
    \includegraphics[width=6.8cm]{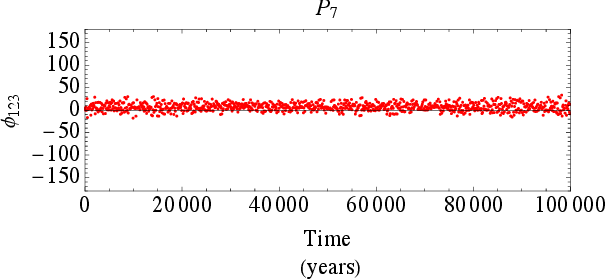}\par 
    \includegraphics[width=6.8cm]{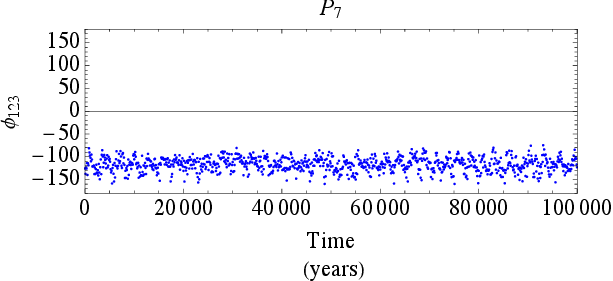}\par
    \end{multicols}
\begin{multicols}{2}
    \includegraphics[width=6.8cm]{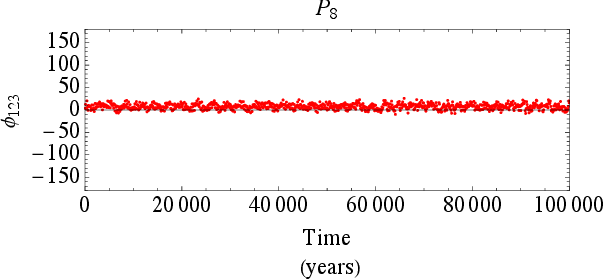}\par
    \includegraphics[width=6.8cm]{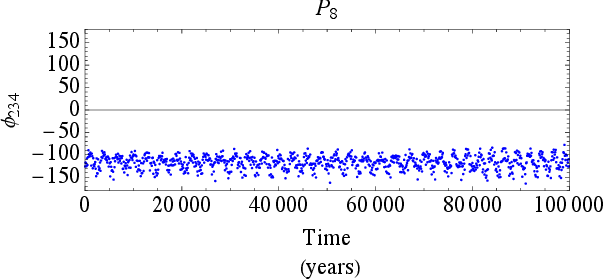}\par
\end{multicols}
\begin{multicols}{2}
    \includegraphics[width=6.8cm]{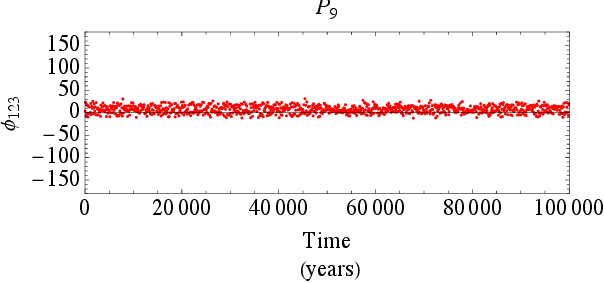}\par
    \includegraphics[width=6.8cm]{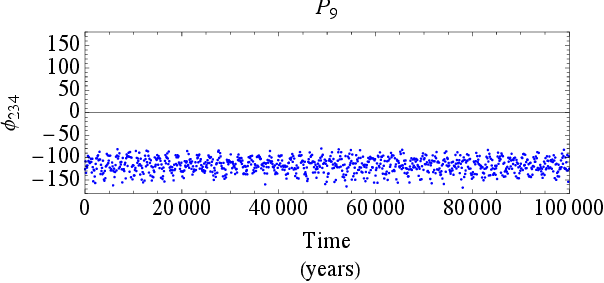}\par
\end{multicols}
\caption{Three--body resonant angles, $\phi_{1 2 3}$ (shown in red) and $\phi_{2 3 4}$ shown in blue,  corresponding to the points 7,8 and 9 that appear in Figure~\ref{A1}. It is clear that small changes in the a--e parameters have changes in these two angles}
\end{figure}

\end{appendices}
\newpage
\clearpage


\bmhead{Acknowledgements}
We thank all the referees for their useful comments and corrections that helped to improve the quality of this article. We thank Prof. Daniel C. Fabrycky for kindly providing initial conditions from his studies of Kepler--223. CEC would like to thank IAChR and JRChR for their helpful discussions and WBRA for her advice and help in the development of this article. CEC especially would like to thank CGrAu for all her advice and kind help in the development of this research.

\bmhead{Author contributions}
All authors contributed equally to this work.

\bmhead{Funding statement}
This research was supported by the Munich Institute for Astro-, Particle and BioPhysics (MIAPbP) which is funded by the Deutsche Forschungsgemeinschaft (DFG, German Research Foundation) under Germany's Excellence Strategy -- EXC 2094 -- 390783311.

\bmhead{Data availability}
The data presented and discussed in this article will be shared on reasonable request to the corresponding author.

\bmhead{Code availability}
Not applicable.

\bmhead{Declarations}

\bmhead{Ethics approval}
Not applicable.

\bmhead{Competing interests}
The authors declare no competing interests.

\bibliography{sn-bibliography}

\end{document}